\documentstyle{mn}
\title{Spacetime modes of relativistic stars}

\author[Nils Andersson, Kostas D. Kokkotas and Bernard F. Schutz]
{ Nils Andersson$^{1}$, Kostas D. Kokkotas$^{1,2}$ 
and Bernard F. Schutz$^{1}$ \\
$^{1}$ Department of Physics and Astronomy,
University of Wales College of Cardiff,
Cardiff CF2 3YB, United Kingdom \\
$^{2}$ Department of Physics, Aristotle University of Thessaloniki,
Thessaloniki 54006, Greece} 

\date{Accepted 1995  (?).
      Received 1995  (?);
      in original form  1995}

\begin{document}

\maketitle

\begin{abstract}
The problem of relativistic stellar pulsations is studied in a somewhat ad hoc 
approximation that ignores all fluid motions. This {\em Inverse
Cowling Approximation} (ICA) is motivated by two observations: 
1) For highly damped ($w$-mode) oscillations the fluid plays very little role. 
2) If the fluid motion is neglected the problem for polar oscillation 
modes becomes similar to that for axial modes. 
Using the ICA we find a polar mode spectrum that has all features of 
the $w$-mode spectrum of the full problem.  
Moreover, in the limit of superdense stars, we find the ICA 
spectrum to be qualitatively similar to that of the axial modes. 
These results clearly show the importance of general relativity for 
the pulsation modes of compact stars, and that there are modes 
whose existence do not depend on motions of the fluid at all; pure ``spacetime'' modes.

\end{abstract}


\section{Introduction}

The pulsations of relativistic stars have attained increasing interest
in the scientific community in the last four years. Recent results have
improved our understanding of both this subject and the theory of general
relativity itself.  Nevertheless, one can easily identify issues that are far 
from well understood at the present time. 

When Thorne and his colleagues established the theory for
relativistic stellar pulsations three decades ago \cite{thorne67,thorne69}, 
the main results
were not that different from the well-known ones of Newtonian theory.
Compact objects where found to oscillate at almost exactly the
frequencies that Newtonian theory predicted. Relativistic
effects only gave rise to a very slow damping of the pulsation.
 The emission of gravitational radiation  implied that the 
oscillation frequencies  were
complex with a relatively tiny imaginary part.

The situation recently changed considerably when a new family of 
oscillation modes for compact stars was found \cite{kojima,kokkotas92}.
 These modes have no relation to the known $p$ and
$g$-modes in Newtonian theory.  Rather, this new family of modes,
the existence of which was suggested by a simple model 
problem \cite{kokkotas86}, comes from coupling
the stellar fluid to the spacetime.  A characteristic property of the new
family is that the fluid motion is very small, and the oscillations 
are rapidly damped.  These modes have been named $w$-modes
(gravitational $w$ave modes), and their existence (together with an
additional part of the spectrum) has been verified by Leins, Nollert and Soffel 
\shortcite{leins93}. 
Moreover, we have recently developed
an accurate numerical code that gives reliable results and also 
reveals the limitations of the specific description of the pulsation problem 
that has been used in all calculations so far \cite{andersson95}.

The oscillations of relativistic 
stars are often described by a 
system of four coupled ODEs \cite{detweiler85}. Two equations correspond 
mainly to the fluid
pulsations and the other two to perturbations of the spacetime.
The two sets of equations couple strongly in the case of highly
relativistic stars, but the coupling also depends on the frequency. 
For slowly damped modes, the equations are in practice decoupled in 
the high and low-frequency limits. These limits correspond to the fluid 
$p$ and $g$-modes, respectively.  The two spacetime ODEs then
hardly affect the structure of the spectrum at all, and a relativistic
generalisation of the Cowling approximation \cite{cowling}, in which 
perturbations of the spacetime itself are ignored,  yields the
correct spectrum [see for example Robe \shortcite{robe68},
McDermott, Van Horn and Scholl \shortcite{mcdermott83}, Finn \shortcite{finn88} 
and Lindblom and Splinter \shortcite{lind90}].

Inversely, as has been shown by Kokkotas and Schutz
\shortcite{kokkotas92}, Leins, Nollert and Soffel \shortcite{leins93} and
Andersson, Kokkotas and Schutz \shortcite{andersson95}, the fluid hardly
pulsates at all in the case of $w$-modes. A simple way to find the
spectrum of such modes might therefore be to omit the equations pertaining 
to the fluid perturbations altogether; what we will from now on 
refer to as the $I$nverse $C$owling $A$pproximation (ICA). 
In the present work, which is intended as a 
complement to studies of the full problem, we examine the stellar
pulsation problem in this approximation.
The major motivation is to clearly show 
the degree of involvement of general relativity. 
The ICA should indicate
which features of the $w$-mode spectrum that are related to the fluid
motion and which are due to the spacetime perturbations. This is
an important step towards a better understanding of the nature of the
$w$-modes and the properties of rapidly damped stellar oscillations.

Chandrasekhar and Ferrari \shortcite{chandra91} recently showed that
axial quasinormal modes could exist for very compact stars. The
possibility of axial modes had previously been discarded since  axial
perturbations do not couple to oscillations of the stellar fluid.
However, Chandrasekhar and Ferrari realized that ``trapped'' modes can
occur when the star is so compact that the surface lies inside the peak
of the familiar Regge-Wheeler potential barrier \cite{chandra83}.  That
is, the axial modes can be understood in terms of a potential well [see
figure 1 in \cite{chandra91}].  Initially, only a few such axial modes
were found and they were all slowly damped. Recent work by Kokkotas
\shortcite{kokkotas94} revealed that there are a large number of highly
damped axial modes as well. This new set of axial modes is in many ways 
similar to the polar $w$-modes. 

A good reason for studying the axial modes is that the spacetime 
perturbations do not couple to those of the fluid, and thus the system of 
equations becomes a very simple one. In fact, the problem is similar to that 
for a perturbed black-hole, although the boundary conditions (now at $r \to
0$) are, of course, different.  If we freeze the fluid motion in the
case of polar perturbations we arrive at a similar problem. Using the
Inverse Cowling Approximation we can consequently study the polar
pulsations in a  way similar to that used for the axial ones.  This
should provide a better understanding of the relation between
the two sets of pulsation modes for relativistic stars. 

Recently, two of us have been involved in a study  
of superdense stars close to the limit of compactness posed by general 
relativity \cite{kojima95}. 
That study suggested that the origin of the axial modes and the polar $w$-modes 
is the same. 
Both sets are ``spacetime'' modes that do not depend on the stellar 
fluid at all for their existence. If that conclusion is correct one would 
expect \underline{all} features of the $w$-modes to be present in the ICA.

\section{The Inverse Cowling Approximation}

In Regge-Wheeler gauge the perturbed metric can be written
\begin{eqnarray}
ds^2 &=& -e^\nu \left(1+r^\ell H_0 e^{i\omega t}Y_{\ell m}\right)dt^2 -
\nonumber \\ &&
 2i\omega r^{\ell+1} H_1 e^{i\omega t}Y_{\ell m} dt dr + 
\nonumber \\&&
e^\lambda \left(1-r^\ell H_0 e^{i\omega t}Y_{\ell m}\right)dr^2 +
\nonumber \\&& 
r^2\left( 1-r^\ell K e^{i\omega t}Y_{\ell m}\right)\left(d\theta^2 +
\sin^2 \theta d\phi^2\right) \ ,
\label{metric}\end{eqnarray}
with $H_0$, $H_1$ and $K$  functions of $r$ only. $Y_{\ell m}$ are the 
standard spherical harmonics and 
$M(r)$ acts as an effective mass inside radius $r$. It is assumed that 
all perturbations depend on time as $\exp(i \omega t)$. 
We also have
\begin{equation}
e^{-\lambda} = 1 - {2 M \over r} \ .
\end{equation}
The metric function $\nu$, the pressure $p$ and the mass $M$ follow from the 
Tolman-Oppenheimer-Volkov equations that determine a stellar 
equilibrium model.
Moreover, Einstein's equations imply the relation \cite{detweiler85}
\begin{eqnarray}
&&\left[ 3M + {(\ell-1)(\ell+2) \over 2} r - 4\pi r^3 \rho \right] H_0 
= \nonumber \\
&&\left[ \omega^2 r^3 e^{-\lambda-\nu} 
- {\ell(\ell+1)\over 2}(M+4\pi r^3 p) \right] H_1 \nonumber \\
&& - \left[  \omega^2 r^3 e^{-\nu} -  {(\ell-1)(\ell+2) \over 2} r \right.
 \nonumber \\
&& \left. - M -4\pi r^3 p - {e^\lambda \over r} 
[ M + 4 \pi r^3 p]^2 \right] K \ .
\end{eqnarray} 
Hence, only two metric functions remain undetermined.

The full pulsation problem consists of four coupled first order ODEs 
[equations (8)-(11) of Detweiler and Lindblom \shortcite{detweiler85}]
These describe the two remaining spacetime variables $H_1$ and $K$ as well 
as two fluid
ones: $V$ and $W$. In this formulation, the ICA  corresponds to $V=W=0$ and
 the perturbation equations for a specific multipole $\ell$ take the 
 following simple form inside the star;
\begin{equation}
r H_1' = e^{\lambda} \left[ H_0 + K \right]
- \left[ \ell + 1 + {{2 M}\over r} e^{\lambda}  + 4 \pi r^2
e^{\lambda}(p-\rho) \right] H_1\ ,
\end{equation}
\begin{equation}
r K' =  H_0 + {\ell(\ell+1)} H_1 - \left[{\ell +1}
- e^\lambda { \left( M+4\pi r^3 p \right) \over r} \right] K \ ,
\end{equation}
where a prime denotes a derivative with respect to $r$ and (3) 
should be used to replace $H_0$. 

In the exterior the perturbation equations simplify to the well-known
Zerilli equation \cite{fackerell71,chandra83}. It is, in fact,
possible to transform equations (4) and (5) into a single second-order
equation but we have found the form of this equation to be rather
complicated and not numerically convenient. Instead, we approach the
above system of equations numerically in the present study. The
procedure used is a very simple one:  The result of one integration
[initiated with the regular solution to (4) and (5)] from the centre of
the star to the surface is matched to an exterior solution calculated
according to the method described by Andersson, Kokkotas and Schutz
\shortcite{andersson95}. The technique used for finding
eigenfrequencies here is identical to that discussed in our previous
paper.

Before we discuss the numerical results obtained in this way a few 
words of caution are in order. We have not defined the ICA in a
gauge-invariant way here. Such a definition does, in fact, not seem possible.
This does not mean that the idea presented here is without merits, however. 
The 
usefulness of an ad hoc approximation such as the ICA is illustrated by the
actual results obtained. Even though such results
should not be used as indication of new physical phenomena we will show in 
the following sections that they provide interesting information. 
We view the present study as a mathematical experiment that 
provides information on the relative importance of the spacetime variables
and the fluid ones. 

\section{ICA modes for polytropes}

We have calculated the quadrupole quasinormal-mode
spectrum in the ICA for the four stellar models of Kokkotas and
Schutz \shortcite{kokkotas92}.  The results are interesting.
First of all, we could not find the $p$ and $g$-mode spectra. This was
certainly expected since these modes are clearly associated with fluid
perturbations which are absent in our approximation \cite{mcdermott83}. 

In Figure 1 we compare the $w$-mode spectrum to the ICA mode spectrum
for model 2 of Kokkotas and Schutz (1992). The similarities between the
two spectra are evident. The highly damped part of the ICA spectrum,
{\em i.e.,} the $w$-modes, is in excellent qualitative agreement with
the results of an analysis of the full problem.  In fact, the real
parts of the frequencies agree surprisingly well. 
The spacing between consecutive modes is roughly 10 \% smaller in
the ICA than in the full problem. This means that the absolute difference
between the real parts for ``corresponding'' modes in the ICA and the
full problem increases drastically with the frequency.
Nevertheless, the relative
difference [$({\rm Re}\ \omega_{n+1} - {\rm Re}\ \omega_{n})/ {\rm Re}\  
\omega_{n+1}$ where $n$ is an integer labelling the modes]
is typically smaller than 10 \% . 
 As for the damping rates,  the ICA imaginary parts are some
20--30 \% smaller than those for the $w$-modes of the full problem. As
the oscillation frequency increases this difference decreases and the
two spectra approach each other.  All characteristic features of the
spectrum, such as the existence of a few overdamped modes with small
real part [what Leins, Nollert and Soffel \shortcite{leins93} refered to as
$w_{II}$-modes] are found where expected.

Even though a discussion of physical effects here must carry a disclaimer 
because of the gauge-dependency of the ICA it is interesting to speculate on
reasons for why the damping of the ICA
modes is slower than that of the $w$-modes of the full problem. A
hand-waving argument is the following: In the full problem the fluid is
also ``radiating'' gravitational waves with the signature of a $w$-mode
(the perturbations are coupled). The fluid thus acts as a kind of 
``gravitational-wave pump'' and it seems plausible that the dissipation rate 
increases. The observed result would follow if the fluid is 
``a more efficient radiator of
gravitational waves'' than the spacetime itself. But does not the evidence 
from oscillations mainly associated with the fluid (the $p$-modes)
indicate the opposite? The damping of these modes due to 
gravitational radiation
is slow, and the fluid would seem to be a poor radiator. This is certainly 
true, but it is important to remember that the origin of the fluid modes and 
the ``spacetime'' modes we discuss in this paper is quite different. For the 
fluid modes the spacetime perturbations are negligible compared to the fluid 
ones, and the modes are slowly
damped because the coupling between matter and gravitational waves is weak.
The situation is different for the $w$-modes, for which the spacetime 
perturbations play the dominant role. The results in Figure 1 suggest 
that the spacetime curvature traps gravitational waves more effectively 
than does the stellar fluid. This seems plausible, but further study of 
this problem
is needed if we are to understand the actual physics involved.
  
\begin{figure}
\caption{The $w$-mode spectrum compared to the ICA spectrum for 
a simple polytropic stellar model. 
Here, the equation of state is $p=100\ {\rm km}^2 \rho^2$, 
and the star has characteristics: 
$\rho_c = 3\times 10^{15}$ g/cm$^3$, $R=8.861$ km, $M=1.266M_\odot$, 
{\em i.e.,} $2M/R=0.422$. 
This corresponds to model 2 of Kokkotas and Schutz (1992).  }
\end{figure}

\section{ICA Modes for Superdense Stars}

Let us now compare the polar modes obtained in the ICA
to  the axial modes. To do this we consider the uniform density model
that was studied by Chandrasekhar and Ferrari \shortcite{chandra91}, 
Kokkotas \shortcite{kokkotas94} and Kojima, Andersson and Kokkotas 
\shortcite{kojima95}.
For this model the surface of the star can be inside the peak of the
curvature potential.  In this way we have a problem that can loosely 
be characterized as a ``potential with a well inside a barrier'', and
one would expect modes with small imaginary part (which come mainly from the
potential well, {\em i.e.,} act as quasi-bound states in quantum language) 
to exist. Furthermore, as was shown by Kokkotas \shortcite{kokkotas94},  
there are modes with high damping. 

Before we proceed to discuss the present results it is necessary to discuss 
the discrepancy between previous results for axial modes. 
In the study of Kokkotas   \shortcite{kokkotas94} the damping rate of 
the modes was found to be roughly half that found by 
Chandrasekhar and Ferrari \shortcite{chandra91}. 
When investigating this issue we found it to be due to a misprint 
in equation (19) of Chandrasekhar and Ferrari \shortcite{chandra91}. 
If the erroneous equation is used in the numerical calculations the 
results of Kokkotas follow. Once the equation is corrected the numerical 
results are in good agreement with those of Chandrasekhar and Ferrari. 
It is, however, important to stress that the results of Kokkotas are 
qualitatively correct. An infinite spectrum of highly damped modes does, 
indeed, exist.
 
Here we have used the method of Andersson, Kokkotas and Schutz
\shortcite{andersson95} to approach the exterior problem and verify 
the qualitative behaviour of
the recent axial-mode results of Kokkotas \shortcite{kokkotas94}.  
It is worth noticing at this point that, although the numerical results of 
Chandrasekhar and Ferrari \shortcite{chandra91}
are correct their method of outward numerical integration
is very sensitive to the choice of endpoint representing ``infinity''. 
Integration from the surface of the star towards infinity is, 
in fact, not very reliable in this kind of problem, 
especially not when one deals with numbers of the order of $10^{-8}$  
as one must to identify the long-lived modes.
Integration towards the surface of the star , as in the method used here, 
is considerably more stable and accurate.

For these ultracompact models we find that the ICA spectrum is quite similar 
to the axial-mode spectrum obtained
by Kokkotas \shortcite{kokkotas94}, see Figure 2.  
The ICA modes are generally slower damped than the axial modes throughout 
the spectrum. This is reminiscent of the result for polar modes of polytropes 
discussed in the previous section.
Our calculations have also unveiled highly damped modes with 
relatively small real parts in both spectra, see Figure 2.
These modes are, in many ways, similar to 
the ``new'' polar modes identified by Leins, Nollert and Soffel 
\shortcite{leins93} and Andersson, Kokkotas and Schutz \shortcite{andersson95}. 
That similar modes exist also for axial perturbations was not known previously, 
but is evident from our Figure 2.

The slowest damped modes for very compact stars should be viewed as modes 
trapped inside the  curvature potential barrier \cite{kojima95}. 
The slight difference in damping between the ICA modes and the axial 
modes in the first part of the spectrum could be due to the difference 
between the
Regge-Wheeler and the Zerilli potential. Although similar, these two 
potentials
are not the same \cite{chandra83} and a difference in the
stellar spectra should be expected at some level.

\begin{figure}
\caption{
The axial and ICA  spectra 
for a very compact uniform density star. This specific example is a star 
for which $R/M = 2.28$ {\em i.e.,} $2M/R = 0.88$. The modes that correspond to 
the ``new'' polar modes identified by Leins, Nollert and Soffel (1993) 
lie above the 
general ``string'' of modes. That such axial modes exist was not known 
previously.
}
\end{figure}

The rapidly damped modes cannot easily be viewed as trapped modes in this 
sense. Rather, they are analogous to the $w$-modes for less compact stellar 
models discussed in the previous section. In fact, it is worth stressing the 
considerable qualitative similarity between the rapidly damped axial modes and 
the polar $w$-modes. 
(compare Figs. 1 and 2). It should also be remembered that the results of 
Kojima, Andersson and Kokkotas \shortcite{kojima95} show that the axial and 
the polar spectra approach each other as the star gets 
increasingly compact. By extending that study to less compact stellar 
models one may hope to shed further light on the relationship between 
the axial modes and the polar $w$-modes. Such work is presently in progress 
\cite{akk95}. At present it seems clear that, although axial perturbations 
do not couple 
to the stellar fluid \cite{thorne67}, rapidly damped axial modes 
{\em should exist also for less compact stellar models}. That is, models for 
which the 
surface of the star lies well outside the peak of the Regge-Wheeler 
potential barrier should support a branch of strongly damped axial modes. 

\section{Concluding Discussion}

In this short paper we have presented results for a new approximation relevant 
to stellar pulsation problems in general relativity. The Inverse Cowling 
Approximation, which neglects perturbations of the stellar fluid, provides a 
useful tool for probing the role of the spacetime degrees of freedom in this 
problem. We have compared 1) the spectrum of highy damped modes for polar 
perturbations (the $w$-modes) to the corresponding ICA modes for polytropic 
neutron star models and 2) the polar ICA modes to the modes for axial 
perturbations of extremely compact uniform density stars.  The results 
are unequivocal: All essential features of the $w$-mode spectra are 
present in the ICA. This is strong support for the idea that the $w$-modes
are ``spacetime'' modes that not rely on the motions of the fluid for their 
existence.

Motivated by the present results, we would argue that this kind of
approximation can be used to further improve our present understanding
of the stellar pulsation problem. In fact, it works much in the
same way as the Cowling approximation does for pulsations mainly
associated with the fluid degrees of freedom. In that case,
calculations are facilitated by neglecting the perturbations of
spacetime \cite{mcdermott83}. When relativistic stars are considered,
that approximation on its own does not make much sense.
It is clear, {\em eg.} from the existence of $w$-modes, that the
gravitational field must be considered as dynamic if all features of
the problem are to be accounted for. So only if the Cowling
approximation is complemented with something like the present 
approximation can one
infer the relevant physics.
 
\subsection*{Acknowledgments}

KDK acknowledges the financial support and the hospitality 
of the Department of Physics and Astronomy,  UWCC, 
without which this work could not have been undertaken. 
NA acknowledges support from SERC. This work was also supported 
by an exchange program from the British Council and the
Greek GSRT.

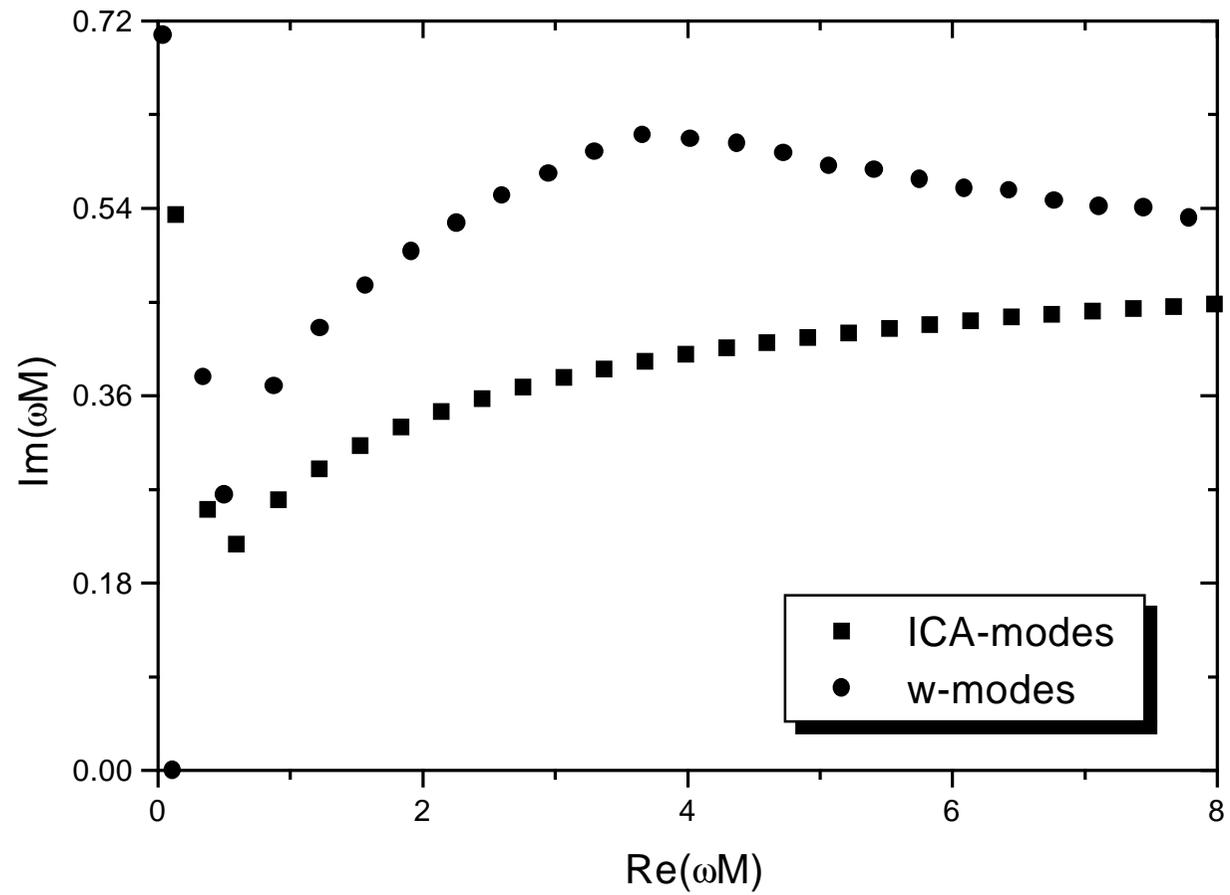
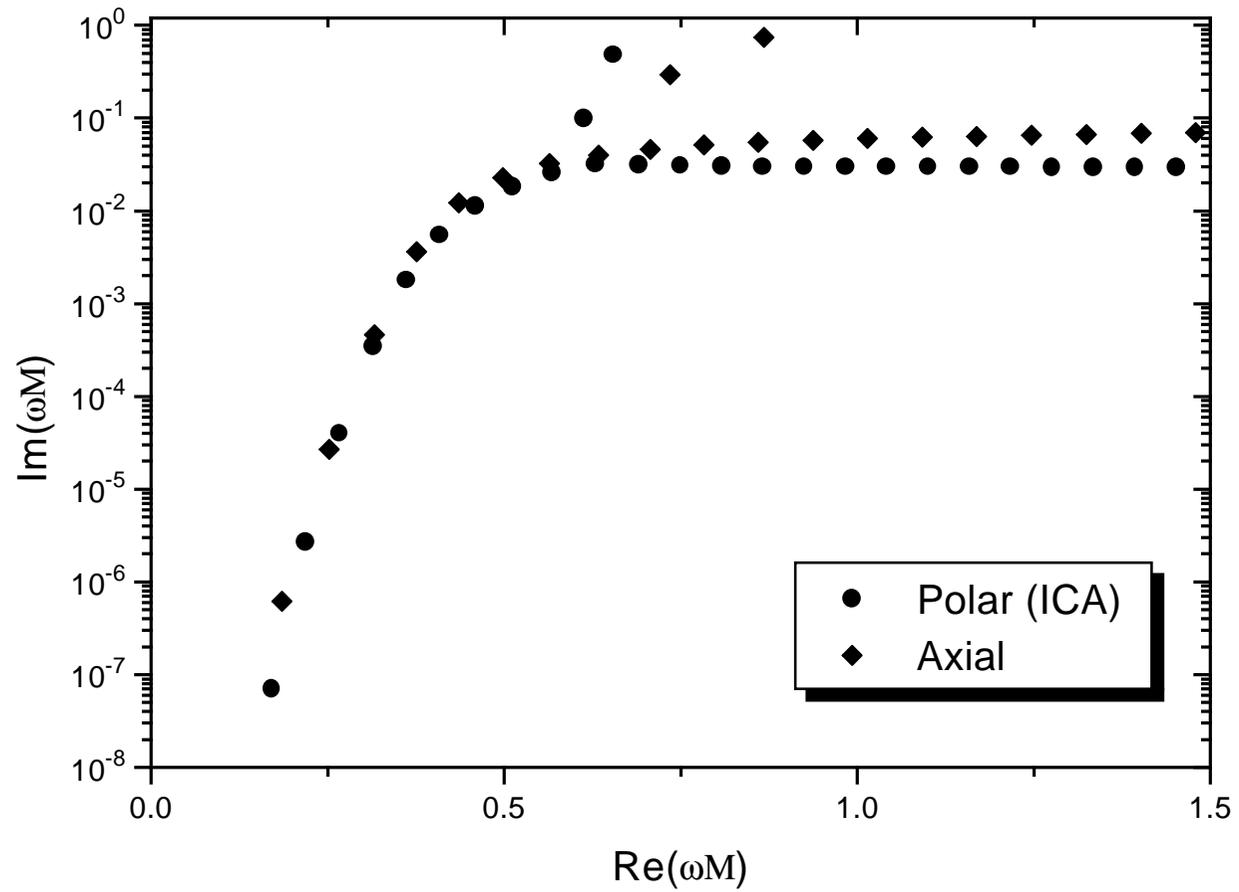
\end{document}